\documentclass[english,aps,pra,10pt,tightenlines,fleqn,twocolumn,superscriptaddress,floatfix]{revtex4-2}
\renewcommand{\bibliography}[1]{}
\pdfoutput=1
\usepackage[utf8]{inputenc}
\usepackage[T1]{fontenc}
\usepackage{amsmath,amssymb,amsfonts}
\usepackage{bm,bbm}
\usepackage{epsfig}
\usepackage{times}
\usepackage{graphics}
\usepackage{booktabs}

\usepackage[dvipsnames]{xcolor}
\definecolor{dblue}{rgb}{0,0.1,.6}
\definecolor{dred}{rgb}{.6,0.1,0}

\newcommand{\dom}[1]   {\textcolor{dred}{#1}}

\usepackage[colorlinks=true,citecolor=dblue,linkcolor=dblue,urlcolor=dblue]{hyperref}
\usepackage[all]{hypcap}

\newcommand{\bra}{\langle}
\newcommand{\ket}{\rangle}
\newcommand{\ud}{\mathrm{d}}

\newcommand{\mc}[1]{\mathcal{#1}}

\newcommand{\A}{\mc{A}}
\newcommand{\B}{\mc{B}}
\newcommand{\C}{\mc{C}}
\newcommand{\E}{\mc{E}}
\renewcommand{\O}{\mc{O}}
\newcommand{\TN}{\text{TN}}
\newcommand{\MPS}{\text{MPS}}
\newcommand{\TTNS}{\text{TTNS}}

\newcommand{\Emph}[1]{\emph{\textbf{#1}}}

\newcommand{\qlab}  {National Quantum Laboratory, University of Maryland, College Park, MD 20742, USA}
\newcommand{\umd}   {Department of Physics, University of Maryland, College Park, MD 20742, USA}
\newcommand{\duke}  {Department of Physics and Duke Quantum Center, Duke University, Durham, NC 27708, USA}

\begin{document}

\title{Cost scaling of MPS and TTNS simulations for 2D and 3D systems with area-law entanglement}
\author{Thomas Barthel}
\affiliation{\qlab}
\affiliation{\umd}
\affiliation{\duke}
\date{October 10, 2026}

\begin{abstract}
Tensor network states are an indispensable tool for the simulation of strongly correlated quantum many-body systems. In recent years, tree tensor network states (TTNS) have been successfully used for two-dimensional systems and to benchmark quantum simulation approaches for condensed matter, nuclear, and particle physics. In comparison to the more traditional approach based on matrix product states (MPS), the graph distance of physical degrees of freedom can be drastically reduced in TTNS. Surprisingly, it turns out that, for large systems in $D>1$ spatial dimensions, MPS simulations of low-energy states are nevertheless more efficient than TTNS simulations. With a focus on $D=2$ and 3, the scaling of computational costs for different boundary conditions is determined under the assumption that the system obeys an entanglement \mbox{(log-)}area law, implying that bond dimensions scale exponentially in the interface area of the associated subsystems.
\end{abstract}

\maketitle

\section{Introduction}
Steve White's density-matrix renormalization group (DMRG) \cite{White1992-11,White1993-10,Schollwoeck2005} revolutionized the simulation of one-dimensional (1D) quantum many-body systems, allowing us to study properties of strongly correlated ground states with very high accuracy -- sometimes approaching machine precision. Although DMRG draws its conceptual inspiration from Wilson's numerical renormalization group \cite{Wilson1975,Bulla2008-80} and quantum information theory, \"{O}stlund and Rommer \cite{Oestlund1995,Rommer1997} showed that DMRG is a variational algorithm on matrix product states (MPS) \cite{Baxter1968-9,Affleck1987-59,Fannes1992-144,Rommer1997,Schollwoeck2011-326}, with the origins of MPS tracing back to the 1940s \cite{Kramers1941-60,Baxter1968-9,Accardi1981,Nightingale1986-33}. Further MPS algorithms allow for the study of finite-temperature states, response functions, non-equilibrium dynamics, and driven-dissipative systems \cite{Jeckelmann2002-66,Vidal2003-10,White2004,Daley2004,Verstraete2004-6,Zwolak2004-93,Sirker2005-71,Feiguin2005-72,White2008-77,Barthel2009-79b,White2009-102,Haegeman2011-107,Karrasch2012-108,Barthel2013-15,Haegeman2016-96,Werner2016-116,Barthel2017_08,Binder2018-98,Paeckel2019-411}.
The entanglement properties of MPS are well-suited for 1D systems. One can prove that 1D states which obey an entanglement-entropy area law, such as ground states of gapped local systems, have faithful MPS approximations \cite{Hastings2007-08,Brandao2013-9,Verstraete2005-5,Barthel2026-9}.

Natural extensions for $D\geq 2$ spatial dimensions are projected entangled-pair states (PEPS) \cite{Niggemann1997-104,Nishino2001-105,Martin-Delgado2001-64,Verstraete2004-7,Verstraete2006-96} and the multi-scale entanglement renormalization ansatz (MERA) \cite{Vidal-2005-12,Vidal2006}. Both can represent states with area-law entanglement \cite{Verstraete2004-7,Verstraete2006-96,Barthel2010-105,Evenbly2014-89a,Ge2016-18} such that, for fixed approximation accuracy, bond dimensions $M$ generally do not need to be increased with system size and grow at most polynomially for target states with log-area-law entanglement. However, while MPS computation costs are only $\Theta(M^3)$, computation costs for 2D PEPS scale as $\Theta(M^{10\dots 12})$ \cite{Jordan2008-101,Orus2009_05} and as $\Theta(M^{16\dots 26})$ for 2D MERA \cite{Cincio2008-100,Evenbly2009-102,Barthel2025-111}, which limits practicable bond dimensions $M$ and the approximation accuracy. Furthermore, while MPS can be optimized and time-evolved with the very efficient and stable DMRG algorithms \cite{White1992-11,White1993-10,McCulloch2008_04,Schollwoeck2011-326,Vidal2003-10,Haegeman2016-96,Vanderstraeten2019-7}, optimization and evolution of MERA \cite{Evenbly2009-79,Rizzi2008-77} are slightly complicated by tensor isometry constraints, and the optimization and evolution of PEPS \cite{Jordan2008-101,Orus2009_05,Vanderstraeten2016-94,Liao2019-9,Czarnik2019-99,Dziarmaga2022-106} are somewhat hampered by the inability to evaluate observables and gradients exactly \cite{Scarpa2020-125,Haferkamp2020-2}.

Tree tensor network states (TTNS) are a middle ground between MPS on the one hand and PEPS or MERA on the other hand. They are closely related to real-space renormalization group schemes \cite{Kadanoff1966-2,Jullien1977-38,Drell1977-16,Martin-Delgado1996-76,Fisher1998-70} and had been considered in analytical work \cite{Affleck1988-115,Fannes1992-66}, before TTNS were optimized implicitly when using DMRG on tree graphs \cite{Otsuka1996-53,Friedman1997-9,Lepetit2000-13} and, finally, described and refined in the tensor-network formulation \cite{Shi2006-74,Tagliacozzo2009-80,Murg2010-82,Nakatani2013-138,Bauernfeind2020-8}. Compared to MPS, the maximal and average graph distances of physical sites are reduced from linear to logarithmic in the system size for TTNS, which is an important advantage. As both are classes of loop-free tensor networks, their associated varieties are closed sets \cite{Barthel2022-112} and the variational optimization is free of barren plateaus \cite{Barthel2023_03,Miao2024-109}. While both can be optimized and evolved efficiently with DMRG and tangent-space algorithms, TTNS incur increased tensor-operation costs of $\Theta(M^{z+1})$, where $z$ is the vertex degree.

In contrast to PEPS and MERA, MPS and TTNS do not match the area-law entanglement structure of typical systems in $D\geq 2$ spatial dimensions \cite{Eisert2008,Latorre2009,Laflorencie2016-646}. Because of their lower tensor-operation costs, algorithmic advantages and limited accessible bond dimensions $M$ on current computers, MPS and TTNS are nevertheless competitive and often preferable for simulations of 2D and 3D systems. Many important works have employed MPS on cylinders and 2D strips \cite{White1998-80,White2007-99,Yan2011-332,Depenbrock2012-109,Stoudenmire2012-3,Jiang2012-8,Jiang2012-86,Cincio2013-110,Zhu2013-110,LeBlanc2015-5,Zheng2017-358,Jiang2019-365,Zaletel2013-110,He2014-112,Zaletel2015-91,Gohlke2017-119,Verresen2019-15,Jiang2021-118,Jiang2022-119,Xu2025_07,Tian2025_08}. In recent years, TTNS have become increasingly popular for the simulation of strongly correlated 2D and 3D systems in the context of condensed matter as well as nuclear and particle physics
\cite{Tagliacozzo2009-80,Murg2010-82,Nakatani2013-138,Bauernfeind2017-7,Gerster2017-96,Macaluso2020-2,Bauernfeind2020-8,Kloss2020-9,Felser2020-10,Cataldi2021-5,Magnifico2021-12,Cao2021-104,Felser2021-126,Jaschke2024-9,Grundner2024-109,Pavesic2025-111,Krinitsin2025-112,Krinitsin2025-134,Pavesic2025_09,Cataldi2025_09}.

This work compares the scaling of computational costs for MPS and TTNS simulations in $D\geq 2$ spatial dimensions with target states that obey area laws or log-area laws in terms of R\'{e}nyi entanglement entropies. As the algorithms for these tensor networks are very similar, the comparison is simply based on the time complexity per optimization or time-evolution step.
Perhaps surprisingly, the results suggest that, in the limit of large system size $L^D$, MPS (with a snake or helical mapping) are more efficient than TTNS with an \emph{exponential} cost separation in the linear system size $L$. At least asymptotically, the increased tensor-operation costs of TTNS outweigh the benefits of smaller graph distances.

Sections~\ref{sec:M-A} and \ref{sec:c_alpha} discuss the relation between MPS and TTNS bond dimensions and R\'{e}nyi entanglement entropies for suitable bipartitions of the system, and Sec.~\ref{sec:contract} reviews tensor-operation costs for both types of tensor networks. 
Based on this, Secs.~\ref{sec:2D}-\ref{sec:hypercube} determine and compare computational costs for the simulation of 2D, 3D, and higher-dimensional systems with MPS and balanced binary TTNS ($z=3$) with rectangular subsystems, assuming entanglement area laws and covering different boundary conditions. Employing geometric isoperimetric inequalities, Sec.~\ref{sec:nonbinary} provides lower bounds for more general TTNS with $z\geq 3$ and demonstrates that vertex degrees $z\geq 4$ are generally not competitive due to increased tensor-operation costs. Finally, Sec.~\ref{sec:logAreaLaw} explains why the exponential cost separation between MPS and TTNS also holds for states with log-area-law entanglement like low-energy states of critical fermionic systems.
In addition to summarizing the results, the conclusion in Sec.~\ref{sec:conclusion} briefly addresses 1D systems, TTNS with physical indices at internal nodes, neglected subexponential cost factors, total versus single-step costs, fast linear algebra, as well as the broader context concerning PEPS and MERA, applications beyond quantum physics, and tensor constraints.

\section{Bond dimensions and subsystem entanglement}\label{sec:M-A}
Let us consider quantum many-body systems on a lattice of $N$ sites and single-site basis states $\{|\sigma_x\ket\,|\,\sigma_x=1,\dotsc,d_x\}$. We will assume the typical scenario, where tensor-network bond dimensions $M_i$ are much larger than the site Hilbert-space dimensions $d_x$, such that computational costs crucially depend on the scaling of bond dimensions with system size.

Extending prior work on MPS \cite{Verstraete2005-5,Schuch2008-100a}, Ref.~\cite{Barthel2026-9} recently showed that bond dimensions for MPS and TTNS approximations $|\psi_\TN\ket$ of a target state $|\psi\ket$ can be bounded by
\begin{equation}\label{eq:Mbound}
	e^{S_{\tilde{\alpha},i}}(1-\delta)^{\frac{\tilde{\alpha}}{\tilde{\alpha}-1}}
	\leq M_i
	\leq e^{S_{\alpha,i}} \left(\frac{\mc{N}_\TN}{\delta}\right)^{\frac{\alpha}{1-\alpha}}.
\end{equation}
Here, $\delta:=\|\psi-\psi_\TN\|^2$ denotes the approximation accuracy of the tensor network state $|\psi_\TN\ket$ with bond dimensions $\{M_i\}$ for the $\mc{N}_\TN=\Theta(N)$ network edges, and $S_{\alpha,i}=\frac{1}{1-\alpha}\ln\sum_\mu\lambda_{i,\mu}^{2\alpha}$ is the $\alpha$-R\'{e}nyi entanglement entropy for the spatial bipartition $\A_i\B_i$ of the system corresponding to cutting edge $i$ of the tensor network with associated Schmidt coefficients $\lambda_{i,1}\geq \lambda_{i,2}\geq\ldots\geq 0$ of $|\psi\ket$. Specifically, Ref.~\cite{Barthel2026-9} shows that any MPS or TTNS approximation with accuracy $\delta$ obeys the lower bound in Eq.~\eqref{eq:Mbound} and that there exist approximations of accuracy $\delta$ with bond dimensions obeying the upper bound in Eq.~\eqref{eq:Mbound}. We need
\begin{equation}\label{eq:alphaRanges}
	\tilde{\alpha}>1\quad\text{and}\quad
	1>\alpha>0
\end{equation}
for the lower and upper bounds, respectively.

The central assumption of this work is that, for a given approximation accuracy, the relevant bond dimensions $M_i$ of edges $i$ in the tensor network scale up to subexponential factors as
\begin{equation}\label{eq:M-A}
	M_i\sim e^{c|\partial \A_i|} =: q^{|\partial \A_i|},
\end{equation}
where $\A_i$ is the smaller of the two subsystems in the spatial bipartition $\A_i\B_i$ of the system, arising from a cut at edge $i$, and $|\partial \A_i|$ is the interface area of that subsystem \footnote{$|\partial\A_i|$ denotes the subsystem interface area, i.e., the $(D-1)$-dimensional measure of the interface separating subsystem $\A_i$ from its complement $\B_i$ within the physical system. For open boundary conditions, the portions of the geometric subsystem surface lying on the external boundary of the system do not contribute. $|\partial\A_i|$ is specified for the corresponding continuum geometries; lattice discretizations are assumed to introduce corrections of order $L^{D-2}$, which do not affect the leading exponential cost scaling and are omitted from the geometric formulas.}.

In Eq.~\eqref{eq:M-A} and the rest of this paper, ``$\sim$'' specifies the leading exponential scaling up to subexponential factors. Specifically, if $L$ denotes a linear (sub)system size and $f(L)$ a positive increasing function in $L$, then
\begin{equation}\label{eq:def-sim}
	\C\sim e^{f(L)}\quad :\Leftrightarrow\quad
	\ln \C= f(L)+o(f(L)),
\end{equation}
where $g(L)=o(f(L))$ means $\lim_{L\to\infty}g(L)/f(L)=0$, i.e., $g$ grows strictly slower than $f$ at large $L$.

According to the bounds \eqref{eq:Mbound}, the scaling \eqref{eq:M-A} is a natural consequence of entanglement area laws, which are typically obeyed by ground and low-energy states of systems where interactions have a finite spatial range or decay sufficiently quickly with distance \cite{Eisert2008,Latorre2009,Laflorencie2016-646}. Intuitively, area laws
\begin{equation}\label{eq:areaLaw}
	S_{\alpha,i} = c|\partial \A_i| + o(|\partial\A_i|)
\end{equation}
arise when all degrees of freedom that can contribute to entanglement between subsystem $\A_i$ and its complement $\B_i$ are located close to the subsystem interface due to an exponential spatial decay of correlations. In critical systems with diverging correlation lengths and power-law decay of correlations, there can be multiplicative logarithmic corrections to the area law. These log-area law systems will be discussed in Sec.~\ref{sec:logAreaLaw}; for now, let us assume the strict area law \eqref{eq:areaLaw}.

In general, the leading area-law term \eqref{eq:areaLaw} can depend on the interface orientation relative to the lattice, while curvature and corners can generate further contributions \cite{Gioev2005,Fradkin2006-97,Casini2007-764,Bueno2015-115,Allais2015-91,Feldman2026-113}. Most cases discussed below involve coordinate-aligned rectangular subsystems $\A_i$. For homogeneous target states with cubic spatial symmetry, all planar subsystem faces then have the same leading entropy density such that Eq.~\eqref{eq:areaLaw} is correct, including subleading contributions from edges and corners. A possible extension to more general subsystem shapes could employ the number $E_i$ of lattice edges crossing the subsystem interface. For example, bounds $c_-E_i\leq S_{1,i}\leq c_+E_i$ with shape-independent constants $c_\pm$ have been established for a class of gapped harmonic lattices \cite{Cramer2006-73}, and one could then combine these with the edge-isoperimetric inequalities of Bollob\'{a}s and Leader \cite{Bollobas1990-4,Bollobas1991-11}. For brevity and lacking more general results in this direction, we do not pursue this route here and employ geometric interface areas $|\partial \A_i|$.

For small subsystem sizes in the lower layers of TTNS -- specifically, length scales smaller than the correlation length or comparable to lattice spacings -- Eq.~\eqref{eq:M-A} may become quite inaccurate. However, we will find that, for large systems, the lower layers of TTNS have negligible contributions to the total computational costs such that these effects can be ignored.

\section{Scaling coefficient in the area law}\label{sec:c_alpha}
The scaling coefficient $c=c_\alpha$ in the area law \eqref{eq:areaLaw} generally depends on the R\'{e}nyi parameter $\alpha$,
\begin{equation}\label{eq:areaLaw-alpha}
	S_{\alpha,i} = c_\alpha |\partial\A_i| + o(|\partial\A_i|).
\end{equation}
Explicit examples of this $\alpha$-dependence are ground states of free scalar fields \cite{Kim2014-733}, $XY$ spin chains \cite{Franchini2008-41}, and string-net models \cite{Flammia2009-103}.
Since the R\'{e}nyi entropy is a non-increasing function of $\alpha$, the leading area-law coefficient obeys
$c_{\tilde{\alpha}}\leq c_1\leq c_\alpha$ for $\tilde{\alpha}>1>\alpha>0$.

Thus, the lower and upper bond-dimension bounds \eqref{eq:Mbound} initially involve different area-law coefficients. One could establish the exponential cost separation of MPS and TTNS simulations by using the upper bound with a suitable $\alpha<1$ for MPS and the lower bound with an $\tilde{\alpha}>1$ for TTNS. However, if $c_\alpha$ is continuous at $\alpha=1$, we can send both R\'{e}nyi parameters to 1, identifying $c=c_1$ in Eq.~\eqref{eq:M-A} at leading exponential order:

Although the R\'{e}nyi entropy of any finite-dimensional density operator is analytic in $\alpha>0$, its large-subsystem asymptotics can exhibit nonanalytic
dependence on $\alpha$, associated with phase transitions of the entanglement Hamiltonian \cite{Metlitski2009-80,Sugino2018-32,Barghathi2026-10,Grover2026_05}. Here we assume that the expansion \eqref{eq:areaLaw-alpha} holds for fixed R\'{e}nyi parameters in a neighborhood of $\alpha=1$ with continuity
\begin{equation}\label{eq:c-alpha-continuity}
	\lim_{\alpha\to 1^-}c_\alpha = c_1 = \lim_{\tilde{\alpha}\to 1^+}c_{\tilde{\alpha}}.
\end{equation}
Analyticity or differentiability is not required.

Let $a_i:=|\partial\A_i|$ such that the bounds \eqref{eq:Mbound} assume the form
\begin{subequations}\label{eq:logM}
\begin{align}
	\ln M_i &\geq c_{\tilde\alpha}\,a_i+o(a_i) +\frac{\tilde\alpha}{\tilde\alpha-1}\ln(1-\delta), \\
	\ln M_i &\leq c_\alpha\, a_i+o(a_i) +\frac{\alpha}{1-\alpha} \ln\frac{\mc{N}_\TN}{\delta}.
\end{align}
\end{subequations}
For the macroscopic interfaces that determine the leading computational costs (as we will find below), one has $a_i\propto L^{D-1}$ for systems of linear size $L$ in $D\geq 2$ spatial dimensions. Provided the total number of sites grows at most polynomially with $L$, we have $\ln\mc{N}_\TN=o(a_i)$ for the number of network edges. Hence, at fixed R\'{e}nyi parameters, the $\delta$-dependent terms in Eqs.~\eqref{eq:logM} are subleading in $a_i$.

For any $\epsilon>0$, continuity \eqref{eq:c-alpha-continuity} allows us to choose fixed
$\tilde\alpha>1$ and $\alpha<1$ sufficiently close to $1$ such that $c_{\tilde\alpha}>c_1-\epsilon$ and $c_\alpha<c_1+\epsilon$.
It follows that, for sufficiently large $L$, every accurate approximation requires $\ln M_i/a_i\geq c_1-2\epsilon$, while approximations exist with $\ln M_i/a_i\leq c_1+2\epsilon$. Since $\epsilon$ is arbitrary, the necessary and achievable leading exponential rates coincide such that
\begin{equation}\label{eq:M-A-precise}
	M_i\sim e^{c_1|\partial\A_i|}.
\end{equation}
This precise version of Eq.~\eqref{eq:M-A} describes asymptotically sufficient bond dimensions with the smallest possible leading exponential rate; unnecessarily enlarged bond dimensions need not satisfy it.
The order of limits is essential. Directly setting $\tilde\alpha=\alpha=1$ would make the bounds \eqref{eq:Mbound} trivial because their exponents diverge.
Instead, we first take the large-system limit and subsequently send the R\'{e}nyi parameters to $1$.

Finally, under the assumption that the relevant interfaces $\partial\A_i$ share a common area-law coefficient $c_1>0$, the subexponential corrections
in Eq.~\eqref{eq:M-A-precise} do not alter the leading cost comparison. Writing $q=e^{c_1}$, we will find costs 
\begin{equation*}
	\C_\MPS\sim q^{3L^{D-1}}\quad \text{and} \quad
	\C_\TTNS\sim q^{m L^{D-1}}
\end{equation*}
for MPS and TTNS with vertex degree $z\geq 3$, respectively, where $m > 3$ depends on the system's boundary conditions and the specific TTNS structure. Thus, $\ln(\C_\TTNS/\C_\MPS)=(m-3)c_1 L^{D-1}+ o(L^{D-1})$, preserving the exponential asymptotic advantage of MPS.

\section{MPS and TTNS tensor-operation costs}\label{sec:contract}
\begin{figure}[t]
	\includegraphics[width=\columnwidth]{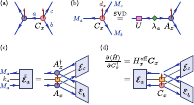}
	\caption{\label{fig:TTNS-operations}
	(a) Part of a TTNS with vertex degree $z=3$, orthogonality center $x$ with edges $a,b,c$, and the corresponding center tensor $C_x$, which has bond dimensions $M_a,M_b,M_c$. Arrows indicate isometric properties of tensors according to the convention of Ref.~\cite{Barthel2026-9}. Vertical lines represent site-basis indices $\sigma_x$ which may be trivial ($d_x=1$). The remaining panels show cost-dominant TTNS operations:
	(b) SVDs which yield Schmidt spectra $\lambda_i$, are used for truncations, and can move the orthogonality center,
	(c) contraction of Hamiltonian environment tensors $\E_i$ and $\tilde{\E}_i$, where the square is a tensor of the Hamiltonian TTN operator with bond dimensions $k_i$,
	(d) contraction of $C_x$ with its effective Hamiltonian $H^\text{eff}_x$, corresponding to the energy gradient $\partial\bra\hat{H}\ket/\partial C_x^\dag$.}
\end{figure}
Throughout this work, we assume conventional dense linear algebra operations \footnote{Throughout this work, computational costs refer to conventional dense linear algebra, including cubic-cost square-matrix multiplication and SVD. However, note that Strassen's algorithm permits subcubic matrix multiplication \cite{Strassen1969-13}, and fast linear-algebra methods also permit numerically stable subcubic decompositions \cite{Demmel2007-108}. Practical speedups have been demonstrated for matrix multiplication and tensor contractions \cite{Huang2016-690,Huang2018-40}, although the benefits depend on matrix dimensions, recursion depth, data movement, and numerical accuracy. Using a fixed number of Strassen recursion levels changes prefactors without changing the asymptotic exponents assumed in this work. Extending the current analysis to asymptotically fast linear algebra makes it necessary to reassess the contractions and decompositions for both MPS and TTNS, but the asymptotic advantage of MPS persists. The numerical cost exponents change, but a simple lower bound from the TTNS tensor sizes already establishes the separation.}.

The most costly operations in MPS optimization and time-evolution algorithms are singular value decompositions (SVD) and effective-Hamiltonian contractions \cite{Schollwoeck2011-326,Vanderstraeten2019-7}. With $M\times M$ matrices, the associated costs scale as
\begin{equation}\label{eq:cost-MPS}
	\Theta(M^3)\quad\text{for MPS},
\end{equation}
where $g(n)=\Theta(f(n))$ means that $g(n)$ grows at the same asymptotic rate as $f(n)$ for large $n$, i.e., there exist constants $k_1,k_2,n_0>0$ such that $0\leq k_1 g(n) \leq f(n) \leq k_2 g(n)$ for all $n \geq n_0$.
Here and in the following, we do not specify costs for sums over Hamiltonian terms or, equivalently, bond indices of the Hamiltonian matrix product operator. These would add a factor that is polynomial in the size of the relevant subsystem $\A_i$, i.e., subexponential terms that are not addressed in this work and captured by the ``$\sim$'' notation introduced in Eq.~\eqref{eq:def-sim}.

For a TTNS with vertex degree $z$, computational costs of single-site algorithms generally scale as
\begin{equation}\label{eq:cost-TTNS-z}
	\Theta(M^{z+1})\quad\text{for TTNS}
\end{equation}
occurring, e.g., in the SVD of a tensor with $z$ bond indices. We will hence focus the analysis on $z=3$, i.e., binary TTNS (Secs.~\ref{sec:2D}-\ref{sec:hypercube}) and address TTNS with larger $z$ more briefly (Sec.~\ref{sec:nonbinary}). As in MPS algorithms, the most costly steps in TTNS algorithms are SVDs, the computation of energy gradients (effective-Hamiltonian contractions), and the propagation of environment tensors (effective Hamiltonians on branches of the tree); see Fig.~\ref{fig:TTNS-operations}. For a $z=3$ TTNS tensor with bond dimensions $M_a,M_b,M_c$, the most costly SVD direction and the dense effective-Hamiltonian and environment contractions have a cost of
\begin{equation}\label{eq:cost-TTNS}
	\Theta\big((M_a)^2 M_b M_c\big)\quad\text{for}\quad
	M_a\geq M_b\geq M_c.
\end{equation}
For SVDs, this follows when considering a matricization with indices $M_b$ and $M_c$ grouped into the matrix column index and $M_a$ associated with the matrix row index. The costs for effective-Hamiltonian contractions and the propagation of environment tensors have the same scaling, because the associated tensor networks, as shown in Figs.~\ref{fig:TTNS-operations}c and \ref{fig:TTNS-operations}d, arise from the same closed tensor network for the energy expectation value by removing one tensor (see Lemma~4 in Ref.~\cite{Barthel2025-111}).
The generalization of Eq.~\eqref{eq:cost-TTNS} for a TTNS node with vertex degree $z$ and bond dimensions $\{M_1,\dotsc,M_z\}$ reads
\begin{equation}\label{eq:cost-TTNS-zM}
	\Theta\Big(\max_j M_j \prod_{j=1}^z M_j\Big)
\end{equation}
in accordance with Eq.~\eqref{eq:cost-TTNS-z}.
The details are discussed in Appendix~\ref{appx:TTNScontractions}.

Note that the tensor operation costs for (naive) two-site TTNS algorithms scale as $\Theta(M^{3z-3})$ and should hence be avoided. However, half renormalization steps make it possible to reduce these costs of two-site TTNS algorithms to $\Theta(M^{z+1})$ \cite{Nakatani2013-138}. Alternatively, convergence problems of single-site algorithms can be circumvented by subspace expansion \cite{Hubig2015-91,McCulloch2024_03}. According to recent algorithmic developments, the exponential scaling \eqref{eq:cost-TTNS-z} of costs in the vertex degree $z$ can be reduced to linear in $z$ by imposing constraints on the tensor CP ranks \cite{Chen2022_05,Chen2024-46}. Nevertheless, for the purpose of this work, we will assume unconstrained tensors as this is currently the common choice in applications.

\section{2D systems}\label{sec:2D}
For the 2D topologies shown in Fig.~\ref{fig:2D}, this section covers MPS and balanced binary TTNS with rectangular subsystems as illustrated in Figs.~\ref{fig:MPS-2D} and \ref{fig:TTNS-2D}.

\subsection{Long cylinder}\label{sec:2D-longCyl}
\begin{figure}[t]
	\includegraphics[width=\columnwidth]{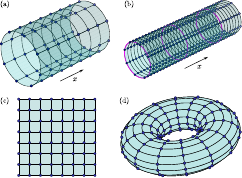}
	\caption{\label{fig:2D}
	2D geometries considered in Sec.~\ref{sec:2D}:
	(a) $L\times L$ lattice on a cylinder with PBC in the $y$ direction,
	(b) long $L_x\times L$ cylinder with $L_x=2^k L$,
	(c) $L\times L$ square with OBC in both directions,
	(d) $L\times L$ lattice on a torus, i.e., PBC in both directions.}
\end{figure}

\begin{figure}[t]
	\includegraphics[width=0.96\columnwidth]{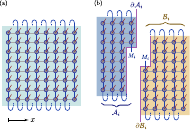}
	\caption{\label{fig:MPS-2D}
	(a) In MPS simulations for 2D systems, we can arrange the MPS sites along a Hamiltonian path that covers the entire lattice, following trails that fully traverse the system in the $y$ direction before progressing in $x$ direction.
	(b) Cutting the MPS tensor network at bond $i$ corresponds to a spatial bipartition into subsystems $\A_i$ and $\B_i$. With Eq.~\eqref{eq:Mbound}, the bond dimensions $M_i$ needed to achieve a certain approximation accuracy can be bounded from above and below by associated R\'{e}nyi entanglement entropies $S_{\alpha,i}$ of the target state. For area-law systems, $S_{\alpha,i}$ is proportional to the interface area $|\partial \A_i|$.}
\end{figure}

In MPS simulations for cylinders of length $L_x$ and circumference $L_y\equiv L$, we can arrange the MPS sites along a Hamiltonian path that covers the entire 2D lattice, following a trail that winds around the cylinder like a snake, fully traversing the system in the $y$ direction before progressing in the $x$ direction. See Figs.~\ref{fig:2D} and \ref{fig:MPS-2D}. One can also work in the limit $L_x\to\infty$ by using infinite MPS \cite{Fannes1992-144,Oestlund1995,Vidal2007-98,Orus2008-78,McCulloch2008_04,Zauner2018-97} with a repeating unit cell of $\propto L$ different tensors.
Cutting any edge $i$ of the MPS splits the cylinder vertically into two segments $\A_i$ and $\B_i$ with an interface of size $|\partial\A_i|=L$ \cite{Note1} such that MPS computation costs scale as 
\begin{equation}\label{eq:2D-MPS}
	\C_\MPS^\text{2D}\stackrel{\eqref{eq:cost-MPS}}{\sim} M_i^3
	\stackrel{\eqref{eq:M-A}}{\sim} q^{3L}.
\end{equation}
\begin{figure*}[t]
	\includegraphics[width=\textwidth]{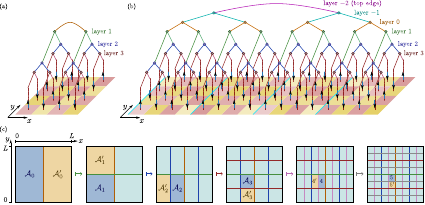}
	\caption{\label{fig:TTNS-2D}
	Simulation of 2D systems using balanced binary TTNS with rectangular subsystems:
	(a) For square geometries, layers split the system alternately in the $x$ and $y$ directions and only the lowest-layer tensors carry physical indices $\sigma_x$.
	(b) For long strips or cylinders, the first layers correspond to a 1D TTNS, splitting the system in the $x$ direction until reaching $L\times L$ segments. 
	(c) A sequence of spatial subsystems, corresponding to bond indices (renormalized sites) after the action of the tensors in layer $n=0,1,2,\dotsc$. We follow a specific sequence, where subsystem $\A_{n-1}$ is split by a tensor of layer $n$ into subsystems $\A_n$ and $\A_n'$ which have interface areas $|\partial \A_n|\geq |\partial \A_n'|$ with respect to their complements. The interface areas can depend on the boundary conditions as specified in Table~\ref{tab:2D}. The scheme works for any 2D lattice, where non-square lattices only require some inessential modifications in the lowest layers.}
\end{figure*}

In simulations with binary TTNS, let us choose $L_x=2^k L$ for simplicity. We then use the first $k-1$ layers of the TTNS to split the cylinder into smaller cylinder segments of length $2^{k-1}L$ in step 1, length $2^{k-2}L$ in step 2, and so on until reaching $L\times L$ cylinders in step $k$ as shown in Fig.~\ref{fig:TTNS-2D}b for $k=2$. In these steps, we have $|\partial\A_i|=2L$ \cite{Note1}, corresponding to two interfaces of size $L$ at the right and left ends of each cylinder segment. The associated computational costs are $\sim q^{8L}$ per tensor operation according to Eqs.~\eqref{eq:M-A} and \eqref{eq:cost-TTNS}.
We then proceed with layers $0,1,2,\dotsc$ splitting each of the $L\times L$ cylinder segments alternately in the $x$ and $y$ directions as indicated in Fig.~\ref{fig:TTNS-2D} with associated subsystem interface areas
$|\partial\A_0| = 2L$,
$|\partial\A_1| = 2L$,
$|\partial\A_2| = \frac{3}{2}L$,
$|\partial\A_3| = L$,
$|\partial\A_4| = \frac{3}{4}L$,
$|\partial\A_5| = \frac{1}{2}L$, etc. So, the subsystem interface area is reduced by a factor $1/2$ every two layers. See also Table~\ref{tab:2D}a.

In an optimization sweep or full time evolution step, we need to pass all tensors of the TTNS. While the number of tensors increases exponentially in the layer index $n$, the subsystem interface areas $|\partial\A_n|$ decrease exponentially in $n$ such that, according to Eq.~\eqref{eq:M-A}, bond dimensions and computational costs per tensor decrease double-exponentially in $n$. Hence, the total cost is dominated by tensor operations for the top layers and we arrive at the scaling
\begin{equation}\label{eq:2D-longCyl-TTNS}
	\C_\TTNS^\text{cyl-long}\stackrel{\eqref{eq:cost-TTNS}}{\sim} M_0^2 M_1^2
	\stackrel{\eqref{eq:M-A}}{\sim} q^{8L},
\end{equation}
for the total cost, which is substantially larger than the MPS cost \eqref{eq:2D-MPS}.
In Eq.~\eqref{eq:2D-longCyl-TTNS}, $M_0,M_1\sim q^{2L}$ refer to the bond dimensions of edges from layer 0 to layer 1 and layer 1 to layer 2, respectively, as indicated in Fig.~\ref{fig:TTNS-2D}b, and the same costs \eqref{eq:2D-longCyl-TTNS} arise for the tensor operations in higher layers.

The described scheme of rectangular $50:50$ splits of subsystems $\A_i\to\A_{i+1}\A'_{i+1}$ minimizes the resulting scaling \eqref{eq:2D-longCyl-TTNS} of the computation costs \footnote{This fact is closely related to relative isoperimetric inequalities for rectangular solids with open or periodic boundary conditions: When bisecting a $D$-dimensional rectangular box $\A_i$ (e.g., a 2D rectangle or 3D cuboid) with open or periodic boundary conditions into two equally sized subsystems $\A_{i+1}$ and $\A'_{i+1}$, the partition that minimizes the area of the interface is a straight cut parallel to the smallest face(s) \cite{Hadwiger1972-76,Ros2005-2,Milman2026-79}. Hence, the resulting subsystems are again rectangular.}.
\begin{table*}[t]
	\centering\renewcommand{\arraystretch}{1.5}
	\begin{tabular}{c c @{\hskip 2.5ex} c}
	  \multicolumn{3}{c}{\textbf{(a)} \ Long cylinder}\\ \addlinespace
	  \toprule
	  Layer & $|\partial\A_n|$ & Cost \\
	  \midrule
	  first $k-1$	& $2L$			& \dom{$q^{8L}$} \\
	  0			& $2L$				& \dom{$q^{8L}$} \\
	  1			& $2L$				& \dom{$q^{8L}$} \\
	  2			& $\frac{3}{2}L$	& $q^{7L}$ \\
	  3			& $L$				& $q^{5L}$ \\
	  4			& $\frac{3}{4}L$	& $q^{\frac{7}{2}L}$ \\
	  5			& $\frac{1}{2}L$	& $q^{\frac{5}{2}L}$ \\
	  \bottomrule
	\end{tabular}
	\qquad \
	\begin{tabular}{c c @{\hskip 2.5ex} c}
	  \multicolumn{3}{c}{\textbf{(b)} \ $L\times L$ cylinder}\\ \addlinespace
	  \toprule
	  Layer & $|\partial\A_n|$ & Cost \\
	  \midrule
	  &&\\
	  0 (top edge)	& $L$				& --\\
	  1			& $\frac{3}{2}L$		& \dom{$q^{\frac{11}{2}L}$} \\
	  2			& $\frac{3}{2}L$, $L$	& \dom{$q^{\frac{11}{2}L}$} \\
	  3			& $L$					& $q^{5L}$ \\
	  4			& $\frac{3}{4}L$		& $q^{\frac{7}{2}L}$ \\
	  5			& $\frac{1}{2}L$		& $q^{\frac{5}{2}L}$ \\
	  \bottomrule
	\end{tabular}
	\qquad \
	\begin{tabular}{c c @{\hskip 2.5ex} c}
	  \multicolumn{3}{c}{\textbf{(c)} \ $L\times L$ square (OBC)}\\ \addlinespace
	  \toprule
	  Layer & $|\partial\A_n|$ & Cost \\
	  \midrule
	  &&\\
	  0 (top edge)	& $L$				& --\\
	  1			& $L$					& $q^{4L}$ \\
	  2			& $\frac{5}{4}L$, $\frac{3}{4}L$	& \dom{$q^{\frac{17}{4}L}$} \\
	  3			& $L$, $\frac{3}{4}L$	& \dom{$q^{\frac{17}{4}L}$} \\
	  4			& $\frac{3}{4}L$		& $q^{\frac{7}{2}L}$ \\
	  5			& $\frac{1}{2}L$		& $q^{\frac{5}{2}L}$ \\
	  \bottomrule
	\end{tabular}
	\qquad \
	\begin{tabular}{c c @{\hskip 2.5ex} c}
	  \multicolumn{3}{c}{\textbf{(d)} \ $L\times L$ torus}\\ \addlinespace
	  \toprule
	  Layer & $|\partial\A_n|$ & Cost \\
	  \midrule
	  &&\\
	  0 (top edge)	& $2L$			& -- \\
	  1			& $2L$				& \dom{$q^{8L}$} \\
	  2			& $\frac{3}{2}L$	& $q^{7L}$ \\
	  3			& $L$				& $q^{5L}$ \\
	  4			& $\frac{3}{4}L$	& $q^{\frac{7}{2}L}$ \\
	  5			& $\frac{1}{2}L$	& $q^{\frac{5}{2}L}$ \\
	  \bottomrule
	\end{tabular}
	\caption{\label{tab:2D}Interface areas $|\partial \A_n|$ (and $|\partial \A'_n|$) for simulations of 2D systems with balanced binary TTNS, where each tensor in layer $n$ splits one subsystem $\A_{n-1}$ of the previous layer into two equal-size subsystems $\A_n$ and $\A'_n$ as illustrated in Fig.~\ref{fig:TTNS-2D}. If $|\partial \A_n|$ and $|\partial \A'_n|$ are different, both are specified, separated by a comma in the second column of each table, and we choose labels such that $|\partial \A_n|\geq|\partial \A'_n|$. Third columns give the associated computational cost per tensor according to Eq.~\eqref{eq:cost-TTNS}, where the three bond dimensions are $\sim q^{|\partial \A_{n-1}|}$, $q^{|\partial \A_n|}$, and $q^{|\partial \A'_n|}$ according to Eq.~\eqref{eq:M-A}. As discussed in Sec.~\ref{sec:2D}, we consider (a) a long $L_x\times L$ cylinder with PBC in the short $y$ direction, (b) an $L\times L$ cylinder, (c) an $L\times L$ square with OBC in both directions, and (d) an $L\times L$ torus.}
\end{table*}

\subsection{\texorpdfstring{$L\times L$}{L\texttimes L} cylinder}
In the comparison to TTNS, long cylinders are somewhat beneficial for MPS as increasing $L_x$ only incurs a polynomial overhead. Let us now consider an $L\times L$ cylinder, again with periodic boundary conditions (PBC) in the $y$ direction. The MPS costs are still given by Eq.~\eqref{eq:2D-MPS}. The top edge of the binary TTNS now corresponds to splitting the cylinder into two $\frac{L}{2}\times L$ segments $\A_0$ and $\A_0'$ as shown in Figs.~\ref{fig:TTNS-2D}a and \ref{fig:TTNS-2D}c. The corresponding interface area is $|\partial\A_0|=|\partial\A_0'|=L$ due to the open boundary conditions (OBC) in $x$ direction. In layer 1, each of the cylinder segments is split horizontally, into two subsystems $\A_1$ and $\A_1'$ of size $\frac{L}{2}\times \frac{L}{2}$ with interface areas $|\partial\A_1|=|\partial\A_1'|=\frac{3}{2}L$, leading to computational costs $\sim q^L(q^{\frac{3}{2}L})^3=q^{\frac{11}{2}L}$ per tensor according to Eqs.~\eqref{eq:M-A} and \eqref{eq:cost-TTNS}. In layer 2, we split $\A_1$ (and similarly all other layer-1 subsystems) further into subsystems $\A_2$ and $\A_2'$ of size $\frac{L}{4}\times \frac{L}{2}$ with interface areas $|\partial\A_2|=\frac{3}{2}L$ and $|\partial\A'_2|=L$, where the asymmetry results from the OBC in $x$ direction. One could try to optimize this splitting but, in the end, the TTNS computational costs scale as
\begin{equation}\label{eq:2D-cyl-TTNS}
	\C_\TTNS^\text{cyl-short}\stackrel{\eqref{eq:cost-TTNS}}{\sim} M_0 M_1^3
	\stackrel{\eqref{eq:M-A}}{\sim} q^{\frac{11}{2} L},
\end{equation}
which is substantially larger than the MPS cost \eqref{eq:2D-MPS}.
See also Table~\ref{tab:2D}b.
In Eq.~\eqref{eq:2D-cyl-TTNS}, $M_0\sim q^{|\partial\A_0|}$ refers to the bond dimension of the top edge and $M_1\sim q^{|\partial\A_1|}$ to the bond dimension of the edges from layer 1 to layer 2 as indicated in Fig.~\ref{fig:TTNS-2D}a.

\subsection{\texorpdfstring{$L\times L$}{L\texttimes L} square}\label{sec:2D-square}
The PBC in $y$ direction of the cylinders that we have considered so far are also in favor of MPS as these PBC do not incur additional costs for MPS. Let us now consider an $L\times L$ square system with OBC in both directions. Following our binary TTNS scheme of splitting subsystems alternately in the $x$ and $y$ directions, we find the associated interface areas and costs per tensor as listed in Table~\ref{tab:2D}c. In layer 1, costs are $\sim q^{4L}$, and one finds $q^{\frac{17}{4}L}=q^{4.25 L}$ for layer 2 which, asymptotically, dominates the total cost. Thus, this construction has
\begin{equation}\label{eq:2D-square-TTNS}
	\C_\TTNS^\text{square}\stackrel{\eqref{eq:cost-TTNS}}{\sim} M_1 M_2' M_2^2
	\stackrel{\eqref{eq:M-A}}{\sim} q^{\frac{17}{4}L} = q^{4.25 L}.
\end{equation}

One might consider optimizing the splittings in layer 2 as, with the mentioned scheme, we get unequal boundary areas $|\partial\A_2|=\frac{5}{4}L$ and $|\partial\A'_2|=\frac{3}{4}L$. However, any construction retaining the initial bisection and subdivision into four equal quadrants still incurs the layer-1 costs $\sim q^{4L}$, which already establishes an exponential separation from the MPS cost \eqref{eq:2D-MPS}.

\newpage
\subsection{\texorpdfstring{$L\times L$}{L\texttimes L} torus}\label{sec:2D-torus}
Finally, consider PBC in both directions for an $L\times L$ system, i.e., a torus.

MPS simulations with PBC in the $x$ direction with the same snake path shown in Fig.~\ref{fig:MPS-2D} are not trivial. When simply using an MPS with OBC -- i.e., not introducing a tensor-network edge that connects the first and last MPS tensors -- bond dimensions get essentially squared compared to the cylinder geometry and computational costs would increase from $\sim q^{3L}$ in Eq.~\eqref{eq:2D-MPS} to $\sim q^{6L}$. Better methods for MPS with PBC and costs $\sim q^{5L}$ and $q^{3L}$ have been introduced in Refs.~\cite{Verstraete2004-4} and \cite{Pippan2010-81}, respectively, but come with some algorithmic complications. If the motivation for $x$-direction PBC in the physical system is to achieve $x$ translation invariance and/or reduce finite-size effects, we can instead more easily work with a long cylinder as discussed in Sec.~\ref{sec:2D-longCyl} such that MPS computational costs are still $\sim q^{3L}$. When sending $L_x\to\infty$, we obtain infinite MPS \cite{Fannes1992-144,Oestlund1995,Vidal2007-98,Orus2008-78,McCulloch2008_04,Zauner2018-97} and $x$ translation invariance.

For binary TTNS on the $L\times L$ torus, we can proceed as above. Splitting subsystems alternately in the $x$ and $y$ directions, we find the associated boundary areas and costs per tensor as listed in Table~\ref{tab:2D}d, resulting in a total TTNS cost of
\begin{equation}\label{eq:2D-torus-TTNS}
	\C_\TTNS^\text{torus}\stackrel{\eqref{eq:cost-TTNS}}{\sim} M_0 M_1^3
	\stackrel{\eqref{eq:M-A}}{\sim} q^{8 L},
\end{equation}
which is again substantially larger than the MPS cost.

\section{3D systems}\label{sec:3D}
For 3D topologies with different boundary conditions, this section covers MPS and balanced binary TTNS with rectangular subsystems.

\subsection{Long cuboid with \texorpdfstring{$yz$}{yz} PBC}\label{sec:3D-longCuboid}
As a first case for 3D, consider cuboids of length $L_x$, width $L_y\equiv L$, and height $L_z\equiv L$ using PBC for the $y$ and $z$ directions. As in 2D, we can arrange the MPS sites along a Hamiltonian path that covers the entire 3D lattice, following a trail that first covers a $yz$ slice (constant or roughly constant $x$, depending on details of the lattice) before progressing in the $x$ direction to cover the next slice, etc. With infinite MPS, one can also work directly in the limit $L_x\to\infty$ \cite{Fannes1992-144,Oestlund1995,Vidal2007-98,Orus2008-78,McCulloch2008_04,Zauner2018-97}.
Cutting any edge $i$ of the MPS splits the cuboid along a $yz$ slice into two segments $\A_i$ and $\B_i$ with an interface of size $|\partial\A_i|= L^2$ \cite{Note1} such that MPS computational costs scale as 
\begin{equation}\label{eq:3D-MPS}
	\C_\MPS^\text{3D}\stackrel{\eqref{eq:cost-MPS}}{\sim} M_i^3
	\stackrel{\eqref{eq:M-A}}{\sim} q^{3L^2}.
\end{equation}

\begin{figure}[t]
	\includegraphics[width=0.95\columnwidth]{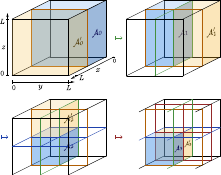}
	\caption{\label{fig:TTNS-3D}
	Simulation of 3D systems using balanced binary TTNS with rectangular subsystems: For cubic geometries, layers split the system cyclically in the $x$, $y$, and $z$ directions. Here, we follow a specific sequence of subsystems, where $\A_{n-1}$ is split by a layer-$n$ tensor into subsystems $\A_n$ and $\A_n'$. The subsystem interface areas are given in Table~\ref{tab:3D} and depend on the boundary conditions.}
\end{figure}

In simulations with binary TTNS, let us choose $L_x=2^k L$. In analogy to Sec.~\ref{sec:2D-longCyl}, the first $k-1$ layers of the TTNS split the cuboid into smaller cuboid segments $\A_n$ of length $2^{k-1}L$ in step 1, length $2^{k-2}L$ in step 2, and so on until reaching $L\times L\times L$ cubes in step $k$. In these steps, we have $|\partial\A_n|= 2L^2$, corresponding to two interfaces of size $L^2$ at the right and left ends of each cuboid segment. The associated computational costs are $\sim q^{8L^2}$ per tensor operation according to Eqs.~\eqref{eq:M-A} and \eqref{eq:cost-TTNS}.
The subsequent layers $0,1,2,\dotsc$ split each of the $L\times L\times L$ cubes cyclically in the $x$, $y$, and $z$ directions into subsystems of size
\begin{align*}
	|\A_0|&=\frac{L}{2}\times L\times L,\\
	|\A_1|&=\frac{L}{2}\times \frac{L}{2}\times L,\\
	|\A_2|&=\frac{L}{2}\times \frac{L}{2}\times \frac{L}{2},\\
	|\A_3|&=\frac{L}{4}\times \frac{L}{2}\times \frac{L}{2},\\
	|\A_4|&=\frac{L}{4}\times \frac{L}{4}\times \frac{L}{2},\\
	|\A_5|&=\frac{L}{4}\times \frac{L}{4}\times \frac{L}{4},\quad \text{and so on}
\end{align*}
as illustrated in Fig.~\ref{fig:TTNS-3D}. The associated subsystem interface areas are
$|\partial\A_0|= 2\ell_y\ell_z= 2L^2$,
$|\partial\A_1|= 2(\ell_y\ell_z+\ell_x\ell_z)=2L^2$,
$|\partial\A_2|= 2(\ell_y\ell_z+\ell_x\ell_z+\ell_x\ell_y)=\frac{3}{2}L^2$,
$|\partial\A_3|= L^2$,
$|\partial\A_4|= \frac{5}{8}L^2$,
$|\partial\A_5|= \frac{3}{8}L^2$, and so on, where interface areas $|\partial\A_n|$ for all $n\geq 2$ are given by the general formula $2(\ell_y\ell_z+\ell_x\ell_z+\ell_x\ell_y)$ with the linear subsystem sizes $\ell_x,\ell_y,\ell_z$.
Hence, the subsystem interface area is reduced by a factor $1/4$ every three layers. See also Table~\ref{tab:3D}a.

As discussed in Sec.~\ref{sec:2D-longCyl}, the total cost is dominated by tensor operations for the top layers and we arrive at the scaling
\begin{equation}\label{eq:3D-longCuboid-TTNS}
	\C_\TTNS^\text{cuboid-$yz$}\stackrel{\eqref{eq:cost-TTNS}}{\sim} M_0^2 M_1^2
	\stackrel{\eqref{eq:M-A}}{\sim} q^{8L^2},
\end{equation}
for the total cost, which is substantially larger than the MPS cost \eqref{eq:3D-MPS}.
\begin{table*}[t]
	\centering\renewcommand{\arraystretch}{1.5}
	\begin{tabular}{c c @{\hskip 2.5ex} c}
	  \multicolumn{3}{c}{\textbf{(a)} \ Long cuboid with $yz$ PBC}\\ \addlinespace
	  \toprule
	  Layer & $|\partial\A_n|$ & Cost \\
	  \midrule
	  first $k-1$	& $2L^2$		& \dom{$q^{8L^2}$} \\
	  0			& $2L^2$			& \dom{$q^{8L^2}$} \\
	  1			& $2L^2$			& \dom{$q^{8L^2}$} \\
	  2			& $\frac{3}{2}L^2$	& $q^{7L^2}$ \\
	  3			& $L^2$				& $q^{5L^2}$ \\
	  4			& $\frac{5}{8}L^2$	& $q^{\frac{13}{4}L^2}$ \\
	  5			& $\frac{3}{8}L^2$	& $q^{2L^2}$ \\
	  \bottomrule
	\end{tabular}
	\qquad \
	\begin{tabular}{c c @{\hskip 2.5ex} c}
	  \multicolumn{3}{c}{\textbf{(b)} \ Cube with $yz$ PBC}\\ \addlinespace
	  \toprule
	  Layer & $|\partial\A_n|$ & Cost \\
	  \midrule
	  &&\\
	  0 (edge)	& $L^2$					& --\\
	  1			& $\frac{3}{2}L^2$			& \dom{$q^{\frac{11}{2}L^2}$} \\
	  2			& $\frac{5}{4}L^2$			& \dom{$q^{\frac{11}{2}L^2}$} \\
	  3			& $L^2$, $\frac{3}{4}L^2$	& $q^{\frac{17}{4}L^2}$ \\
	  4			& $\frac{5}{8}L^2$			& $q^{\frac{13}{4}L^2}$ \\
	  5			& $\frac{3}{8}L^2$			& $q^{2L^2}$ \\
	  \bottomrule
	\end{tabular}
	\qquad \
	\begin{tabular}{c c @{\hskip 2.5ex} c}
	  \multicolumn{3}{c}{\textbf{(c)} \ Cube with $xyz$ OBC}\\ \addlinespace
	  \toprule
	  Layer & $|\partial\A_n|$ & Cost \\
	  \midrule
	  &&\\
	  0 (edge)	& $L^2$								& --\\
	  1			& $L^2$									& \dom{$q^{4L^2}$} \\
	  2			& $\frac{3}{4}L^2$						& $q^{\frac{7}{2}L^2}$ \\
	  3			& $\frac{3}{4}L^2$,	$\frac{1}{2}L^2$	& $q^{\frac{11}{4}L^2}$ \\
	  4			& $\frac{9}{16}L^2$, $\frac{7}{16}L^2$	& $q^{\frac{5}{2}L^2}$ \\
	  5			& $\frac{3}{8}L^2$,  $\frac{5}{16}L^2$	& $q^{\frac{29}{16}L^2}$ \\
	  \bottomrule
	\end{tabular}
	\qquad \
	\begin{tabular}{c c @{\hskip 2.5ex} c}
	  \multicolumn{3}{c}{\textbf{(d)} \ Cube with $xyz$ PBC}\\ \addlinespace
	  \toprule
	  Layer & $|\partial\A_n|$ & Cost \\
	  \midrule
	  &&\\
	  0 (edge)	& $2L^2$			& -- \\
	  1			& $2L^2$			& \dom{$q^{8L^2}$} \\
	  2			& $\frac{3}{2}L^2$	& $q^{7L^2}$ \\
	  3			& $L^2$				& $q^{5L^2}$ \\
	  4			& $\frac{5}{8}L^2$	& $q^{\frac{13}{4}L^2}$ \\
	  5			& $\frac{3}{8}L^2$	& $q^{2L^2}$ \\
	  \bottomrule
	\end{tabular}
	\caption{\label{tab:3D}Interface areas $|\partial \A_n|$ (and $|\partial \A'_n|$) for simulations of 3D systems with balanced binary TTNS, where each tensor in layer $n$ splits one subsystem $\A_{n-1}$ of the previous layer into two equal-size subsystems $\A_n$ and $\A'_n$ as illustrated in Fig.~\ref{fig:TTNS-3D}. Third columns give the associated computational cost per tensor according to Eq.~\eqref{eq:cost-TTNS}. As discussed in Sec.~\ref{sec:3D}, we consider (a) a long $L_x\times L\times L$ cuboid with PBC in the short $yz$ directions, (b) an $L^3$ cube with $yz$ PBC, (c) an $L^3$ cube with OBC in all directions, and (d) an $L^3$ cube with PBC in all directions.}
\end{table*}

\subsection{\texorpdfstring{$L^3$ cube with $yz$ PBC}{L\textthreesuperior cube with yz PBC}}
Consider now an $L\times L\times L$ cube, again with PBC in the $y$ and $z$ directions. The MPS costs are still given by Eq.~\eqref{eq:3D-MPS}. Again, we split cyclically in the $x$, $y$, and $z$ directions such that the resulting subsystems $\A_n$ for layer $n=0,1,2,\dotsc$ are shaped as specified in Sec.~\ref{sec:3D-longCuboid} and Fig.~\ref{fig:TTNS-3D}, but the boundary areas $|\partial\A_n|$ have changed due to the OBC in $x$ direction. For example, $|\partial\A_0|=\ell_y\ell_z=L^2$, $|\partial\A_1|=\ell_y\ell_z+2\ell_x\ell_z=\frac{3}{2}L^2$, and  $|\partial\A_2|=\ell_y\ell_z+2(\ell_x\ell_z+\ell_x\ell_y)=\frac{5}{4}L^2$. In layer 3, we split $\A_2$ (and similarly all other layer-2 subsystems) further into subsystems $\A_3$ and $\A_3'$ of size $\frac{L}{4}\times \frac{L}{2}\times \frac{L}{2}$ with interface areas $|\partial\A_3|=2(\ell_y\ell_z+\ell_x\ell_z+\ell_x\ell_y)=L^2$ and $|\partial\A'_3|=\ell_y\ell_z+2(\ell_x\ell_z+\ell_x\ell_y)=\frac{3}{4}L^2$ with the asymmetry resulting from the OBC in $x$ direction. One could try to optimize this splitting but, in the end, the TTNS computation costs scale as
\begin{equation}\label{eq:3D-cube-yz-TTNS}
	\C_\TTNS^\text{cube-$yz$}\stackrel{\eqref{eq:cost-TTNS}}{\sim} M_0 M_1^3
	\stackrel{\eqref{eq:M-A}}{\sim} q^{\frac{11}{2} L^2},
\end{equation}
which is substantially larger than the MPS cost \eqref{eq:3D-MPS}.
See also Table~\ref{tab:3D}b.

\subsection{\texorpdfstring{$L^3$ cube with $xyz$ OBC}{L\textthreesuperior cube with xyz OBC}}\label{sec:3D-cube}
PBC in the $yz$ directions do not affect the asymptotic cost for MPS, but they do increase the cost for TTNS. To see this explicitly, let us now consider an $L\times L\times L$ cube with OBC in all directions. Following our TTNS scheme of splitting subsystems cyclically in the $x$, $y$, and $z$ directions, we find the associated boundary areas and costs per tensor as listed in Table~\ref{tab:3D}c. Asymptotically, the total TTNS cost is dominated by the layer-1 cost
\begin{equation}\label{eq:3D-cube-TTNS}
	\C_\TTNS^\text{cube}\stackrel{\eqref{eq:cost-TTNS}}{\sim} M_0 M_1^3
	\stackrel{\eqref{eq:M-A}}{\sim} q^{4L^2}.
\end{equation}

\subsection{\texorpdfstring{$L^3$ cube with $xyz$ PBC}{L\textthreesuperior cube with xyz PBC}}\label{sec:3D-cube-xyz}
Finally, consider PBC in all three directions for an $L\times L\times L$ cube.

Following the discussion in Sec.~\ref{sec:2D-torus}, the algorithm from Ref.~\cite{Pippan2010-81} achieves MPS costs $\sim q^{3L^2}$ [Eq.~\eqref{eq:3D-MPS}]. If the motivation for $x$-direction PBC in the physical system is to achieve $x$ translation invariance and/or reduce finite-size effects, one can instead more easily use OBC in $x$ direction and work with a long cuboid (Sec.~\ref{sec:2D-longCyl}) or even send $L_x\to\infty$ using infinite MPS \cite{Fannes1992-144,Oestlund1995,Vidal2007-98,Orus2008-78,McCulloch2008_04,Zauner2018-97} at cost $\sim q^{3L^2}$.

For TTNS, we can proceed as above. Splitting subsystems cyclically in the $x$, $y$, and $z$ directions, we find the associated boundary areas and costs per tensor as listed in Table~\ref{tab:3D}d, resulting in a total TTNS cost of
\begin{equation}\label{eq:3D-cube-xyz-TTNS}
	\C_\TTNS^\text{cube-$xyz$}\stackrel{\eqref{eq:cost-TTNS}}{\sim} M_0 M_1^3
	\stackrel{\eqref{eq:M-A}}{\sim} q^{8 L^2},
\end{equation}
which is again substantially larger than the MPS cost.

\section{\texorpdfstring{$L^D$ hypercube with PBC in $D\geq 2$ dimensions}{L\textasciicircum D hypercube with PBC in D>=2 dimensions}}\label{sec:hypercube}
More generally, let us consider a hypercube of side length $L$ in $D\geq 2$ spatial dimensions with PBC in all directions (a flat torus).

For MPS, we can use the approach from Secs.~\ref{sec:2D-torus} and \ref{sec:3D-cube-xyz}. Either employing the method from Ref.~\cite{Pippan2010-81}, or sending $L_x\to\infty$ and using OBC in the $x$ direction instead of PBC, we achieve the cost scaling
\begin{equation}\label{eq:hypercube-MPS}
	\C_\MPS^\text{hypercube}\stackrel{\eqref{eq:cost-MPS}}{\sim} M_i^3
	\stackrel{\eqref{eq:M-A}}{\sim} q^{3L^{D-1}}.
\end{equation}

For balanced binary TTNS, both the original hypercube geometry and the modified geometry with $L_x\to \infty$ and $x$ OBC lead to the cost scaling
\begin{equation}\label{eq:hypercube-TTNS}
	\C_\TTNS^\text{hypercube}\stackrel{\eqref{eq:cost-TTNS}}{\sim} M_0 M_1^3
	\stackrel{\eqref{eq:M-A}}{\sim} q^{8L^{D-1}}.
\end{equation}
With subsystem $\A_0$ haveing shape $\frac{L}{2}\times L^{D-1}$, we then split cyclically in every spatial dimension when going through TTNS layers $n=1,2,3\dotsc$. With $|\partial\A_0|=2L^{D-1}$ and $|\partial\A_1|=|\partial\A'_1|=2L^{D-1}$, we obtain Eq.~\eqref{eq:hypercube-TTNS}.

\section{More general TTNS with \texorpdfstring{$z\geq 3$}{z>=3}}\label{sec:nonbinary}
\begin{figure}[t]
	\includegraphics[width=0.95\columnwidth]{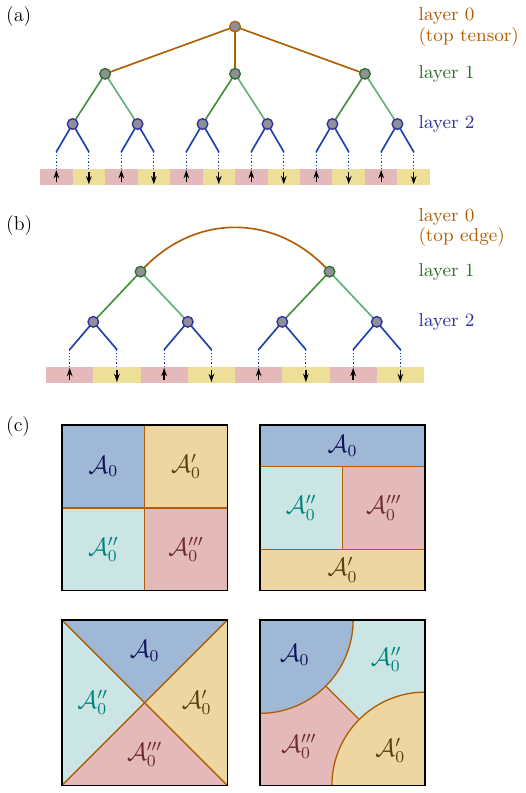}
	\caption{\label{fig:TTNS-roots-partitions}
	Section~\ref{sec:nonbinary} addresses more general classes of TTNS with vertex degrees $z\geq 3$.
	(a) They can either have a top tensor that splits the system into $z$ equal-volume subsystems, or
	(b) a top edge that bisects the system and is followed by two tensors that each split one half of the system into $z-1$ equal-volume subsystems. The rest of the TTNS can be arbitrary.
	(c) As illustrated in this panel for $z=4$ and a 2D square system, the top tensor can decompose the system into $z$ equally shaped rectangular subsystems (top left) similar to the balanced binary TTNS analyzed in Secs.~\ref{sec:2D}-\ref{sec:hypercube}, rectangular subsystems of different shape (top right), or arbitrarily shaped subsystems (bottom), which we still constrain to have equal volume.}
\end{figure}
Sections~\ref{sec:2D}-\ref{sec:hypercube} covered MPS and balanced binary TTNS ($z=3$) with rectangular subsystems for different system geometries. The exponential increase of tensor-operation costs \eqref{eq:cost-TTNS-z} with the vertex degree $z$ suggests that balanced binary TTNS should generally be the most efficient choice. To substantiate this claim/intuition and to establish the exponential disadvantage for TTNS more generally, let us consider a considerably more general class of TTNS with vertex degrees $z\geq 3$ in this section. We will again assume Eq.~\eqref{eq:M-A-precise} with a common area-law coefficient $c_1>0$ for the relevant interfaces and set $q:=e^{c_1}$.

For the top layer of the TTNS, we consider two arrangements:
\begin{enumerate}
\itemsep=0pt
	\item[(a)] a single top tensor that divides the system into $z$ equal-volume subsystems, and
	\item[(b)] a top edge that first bisects the system with both endpoints subsequently dividing their halves into $z-1$ equal-volume subsystems.
\end{enumerate}
For the lower bounds on TTNS costs discussed in the following, we will allow for arbitrary subsystem shapes. Only the splittings in the upper layers are assumed to have vertex degree $z$ and be balanced, i.e., to produce equal-volume subsystems. The remaining tree is unrestricted. For case (a), this only concerns the top tensor. For case (b), it concerns the top edge and the tensors at its end points. See Fig.~\ref{fig:TTNS-roots-partitions}.

For a TTNS vertex $x$ with vertex degree $z$, let $|\partial\A_1|,\dotsc,|\partial\A_z|$ denote the interface areas \cite{Note1} associated with each edge incident to $x$. According to Eq.~\eqref{eq:cost-TTNS-zM}, conventional dense tensor operations \cite{Note2} discussed in Appendix~\ref{appx:TTNScontractions-z} have leading cost
\begin{subequations}\label{eq:cost-TTNS-zB}
\begin{align}
	\C_x&\sim q^{B_x}\quad\text{with}\\
	B_x&:=\sum_{j=1}^z |\partial\A_j|+\max_{1\leq j\leq z}|\partial\A_j|,
\end{align}
\end{subequations}
i.e., the relevant quantity $B_x$ is the sum of the incident interface areas plus the largest one.

\subsection{Degree-four TTNS for an \texorpdfstring{$L\times L$}{L\texttimes L} square}\label{sec:nonbinary-z4}
(a) For $z=4$, let a top tensor divide the $L\times L$ square with OBC into four equal-area subsystems. The sharp relative isoperimetric
inequality for a square is \cite{Brezis2021-359}
\begin{equation}\label{eq:square-isoperi}
	|\partial\A_j|\geq \min\{1,\sqrt{\pi v}\}\,L\quad\text{for}\quad
	v:=\frac{|\A_j|}{L^2}\leq\frac{1}{2},
\end{equation}
where only the interface with $\A_j$'s complement $\B_j$ on the square is counted \cite{Note1}. Applying this bound to the four equal-area root
branches with $v=1/4$ gives
\begin{equation}\label{eq:nonbinary-z4-B0}
	B_0\geq\frac{5\sqrt\pi}{2}\,L \approx 4.43L.
\end{equation}
When choosing a splitting into four $\frac{L}{2}\times\frac{L}{2}$ squares at the top, one finds $B_0=5L$.

(b) For the case with an initial top edge, a straight bisection gives two $\frac{L}{2}\times L$ rectangles with interface length $L$. Dividing each rectangle
into three $\frac{L}{2}\times \frac{L}{3}$ strips gives child interface lengths $(\frac{5L}{6},\frac{4L}{3},\frac{5L}{6})$, so that the first degree-four vertices have
\begin{equation}\label{eq:nonbinary-z4-B1-rect}
	B_1\stackrel{\eqref{eq:cost-TTNS-zB}}{=} L+\left(\frac{5}{6}+\frac{4}{3}+\frac{5}{6}\right)L+\frac{4}{3}L \approx 5.33 L,
\end{equation}
again exceeding the exponent $4.25L$ of the balanced binary TTNS in Eq.~\eqref{eq:2D-square-TTNS}.
For unrestricted shapes, the top edge has $v=1/2$ and the children have $v=1/6$ such that the isoperimetric inequality~\eqref{eq:square-isoperi} gives
\begin{equation}\label{eq:nonbinary-z4-B1}
	B_1\geq\left(2+3\sqrt{\pi/6}\right)L\approx 4.17L.
\end{equation}

So, in all $z=4$ scenarios described here, the leading TTNS cost exponent hence exceeds the MPS exponent $3L$ in Eq.~\eqref{eq:2D-MPS}. Furthermore, all bounds except that in Eq.~\eqref{eq:nonbinary-z4-B1} establish a disadvantage of $z=4$ relative to the exponent $\frac{17}{4}\,L = 4.25L$ of the balanced binary TTNS in Eq.~\eqref{eq:2D-square-TTNS}. Of course, these lower bounds for $z=4$ need not be attainable.

\subsection{Degree-five TTNS for an \texorpdfstring{$L\times L$}{L\texttimes L} square}\label{sec:nonbinary-z5}
(a) For $z=5$, let a top tensor split the square into five equal-area subsystems; each branch has volume fraction $v=1/5$. The isoperimetric inequality \eqref{eq:square-isoperi} then gives $|\partial\A_j|\geq \min\left\{1,\sqrt{{\pi}/{5}}\right\}\,L=\sqrt{{\pi}/{5}}\,L$. Therefore,
\begin{equation}\label{eq:nonbinary-z5-B0}
	B_0\geq 6\sqrt{{\pi}/{5}}\,L \approx 4.76L.
\end{equation}

(b) If a top edge first bisects the square, we can divide each resulting $\frac{L}{2}\times L$ rectangle into four $\frac{L}{4}\times \frac{L}{2}$ rectangles. The parent interface has length $L$, while the child interfaces have lengths $(\frac{3L}{4},\frac{5L}{4},\frac{5L}{4},\frac{3L}{4})$. Hence, the first degree-five vertices have
\begin{equation}\label{eq:nonbinary-z5-B1-rect}
	B_1=L+\left(\frac{3}{4}+\frac{5}{4}+\frac{5}{4}+\frac{3}{4}\right)L+\frac{5}{4}L=\frac{25}{4}L.
\end{equation}
For unrestricted shapes, the top edge has $v=1/2$ and the children have $v=1/8$ such that the isoperimetric inequality~\eqref{eq:square-isoperi} gives
\begin{equation}\label{eq:nonbinary-z5-B1}
	B_1\geq\left(2+4\sqrt{\pi/8}\right)L\approx 4.51L.
\end{equation}

So, in all $z=5$ scenarios described here, the leading TTNS cost exponent exceeds the MPS exponent $3L$, the exponent $4.25L$ of the balanced binary TTNS, and the corresponding exponents or exponent bounds of the $z=4$ TTNS from Sec.~\ref{sec:nonbinary-z4}.

\subsection{General-\texorpdfstring{$z$}{z} TTNS for hypercubes with OBC}\label{sec:nonbinary-obc}
Let us consider a hypercube of side length $L$ in $D\geq 2$ spatial dimensions with OBC in all directions and TTNS with vertex degree $z\geq 3$. With $\Sigma:=L^{D-1}$, a geometric isoperimetric inequality valid for arbitrary subsystem shapes is
\begin{equation}\label{eq:cube-isoperi}
	|\partial\A_j|\geq 4v(1-v)\Sigma\quad\text{for}\quad v:=\frac{|\A_j|}{L^D},
\end{equation}
as established for polyhedral sets by Hadwiger \cite{Hadwiger1972-76} and for general measurable sets by Ambrosio et al.~\cite{Ambrosio2016-69}.

(a.1) Let a top tensor split the system into $z$ equal-volume subsystems. With child volume fraction $v=1/z$, Eqs.~\eqref{eq:cost-TTNS-zB} and \eqref{eq:cube-isoperi} give
\begin{equation}\label{sec:nonbinary-obc-a1}
	B_0\geq (z+1)\,\frac{4}{z}\left(1-\frac{1}{z}\right)\Sigma
	=4\left(1-\frac{1}{z^2}\right)\Sigma.
\end{equation}

(b.1) For the case with a top edge, the initial bisection has an interface area of at least $\Sigma$. Each of the $b:=z-1$
children of either endpoint has volume fraction $v=1/(2b)$. Using the parent interface both in the sum and as a lower bound on the
maximum in Eq.~\eqref{eq:cost-TTNS-zB}, we have
\begin{equation}\label{sec:nonbinary-obc-b1}
	B_1\geq 2\Sigma+ b\,\frac{4}{2b}\left(1-\frac{1}{2b}\right)\Sigma
	=\left(4-\frac{1}{z-1}\right)\Sigma.
\end{equation}

The lower bounds \eqref{sec:nonbinary-obc-a1} and \eqref{sec:nonbinary-obc-b1} for TTNS with arbitrary subsystem shapes exceed the MPS cost exponent $3\Sigma$ but, for $D=2,3$, they are slightly lower than the results \eqref{eq:2D-square-TTNS} and \eqref{eq:3D-cube-TTNS} for the balanced binary TTNS.

Of course, the bounds \eqref{sec:nonbinary-obc-a1} and \eqref{sec:nonbinary-obc-b1} need not be attainable. Stronger bounds relative to the costs $\sim q^{4.25L}$ and $q^{4L^2}$ of the balanced binary TTNS from Secs.~\ref{sec:2D-square} and \ref{sec:3D-cube} follow for coordinate-aligned rectangular subsystems:
Suppose such a subsystem spans the full hypercube in $D-k$ directions and has side lengths $L_1,\dotsc,L_k<L$ in the remaining directions. It has the volume
fraction $v=\prod_{\mu=1}^k(L_\mu/L)$. In every nonspanning direction, at least one face separates the subsystem from its complement. Thus,
\begin{equation}\label{sec:nonbinary-obc-rect}
	|\partial\A_j| \geq vL^D\sum_{\mu=1}^k\frac1{L_\mu} \geq k\, v^{1-\frac{1}{k}}\,\Sigma.
\end{equation}
The first inequality becomes an equality when the rectangle touches an external face in every nonspanning direction, and the second step follows from the inequality of arithmetic and geometric means. 

Let us define the strictly increasing function
\begin{equation}
	h(u):=\min_{k\in\{1,\dotsc,D\}} k\, u^{1/k}.
\end{equation}
For $D\geq 2$ dimensions, it has the values $h(3)=3$, $h(4)=4$, $h(5)=2\sqrt{5}$, $h(6)=2\sqrt{6}$, and $h(8)=4\sqrt{2}$. It also obeys $h(u)\geq 4$ for all $u\geq 4$ with the strict inequality $h(u)>4$ for $u>4$, and $\frac{u+1}u\, h(u)$ is strictly increasing for $u\geq 3$ 
\footnote{Each branch $k\,u^{1/k}$ is strictly increasing. On a minimizing branch, $\frac{\ud}{\ud u}[(1+1/u)\,k\,u^{1/k}]=u^{1/k-2}(u+1-k)>0$ because the
minimizing integer satisfies $k\leq\lceil\ln u\rceil < u+1$. Continuity at branch transitions then establishes the stated monotonicity.}.

(a.2) For a top tensor that splits the system into $z$ coordinate-aligned rectangular subsystems with volume ratio $v=1/z$, we obtain
\begin{equation}\label{sec:nonbinary-obc-a2}
	B_0\geq \frac{z+1}{z}\,h(z)\Sigma\
	\begin{cases}
	= 4\Sigma		&\text{for}\ z=3,\\
	\geq 5\Sigma	&\text{for}\ z\geq 4.\\
	\end{cases}
\end{equation}

(b.2) With an initial top edge bisecting the system and the following tensors splitting each half into $b=z-1$ coordinate-aligned rectangular subsystems with volume ratio $v=1/(2b)$, we have
\begin{equation}\label{sec:nonbinary-obc-b2}
	B_1\geq \frac{1}{2}\big(4+h(2z-2)\big)\Sigma
\begin{cases}
	=4\Sigma				&\text{for}\ z=3,\\
	\geq (2+\sqrt{6})\Sigma	&\text{for}\ z\geq 4.\\
	\end{cases}
\end{equation}

For $z\geq 4$, both rectangular bounds \eqref{sec:nonbinary-obc-a2} and \eqref{sec:nonbinary-obc-b2} do not only exceed the MPS cost exponent $3\Sigma$ but, for $D=2,3$, also the exponents $4L\dots 4.25L$ and $4L^2$ for the open square and cube [Eqs.~\eqref{eq:2D-square-TTNS} and \eqref{eq:3D-cube-TTNS}]. Note that $2+\sqrt{6}\approx 4.45$.

\subsection{General-\texorpdfstring{$z$}{z} TTNS for hypercubes with PBC}\label{sec:nonbinary-pbc}
For a hypercube of side length $L$ with PBC in all directions (a flat torus), the geometric isoperimetric inequality \eqref{eq:cube-isoperi} for OBC can be extended to PBC by reflection \cite{Ros2005-2,Milman2026-79}, giving
\begin{equation}\label{eq:torus-isoperi}
	|\partial\A_j|\geq 8v(1-v)\Sigma\quad\text{for}\quad v:=\frac{|\A_j|}{L^D}.
\end{equation}
Additionally, in adaptation of Eq.~\eqref{sec:nonbinary-obc-rect}, we now have 
\begin{equation}
	|\partial\A_j|=2vL^D\sum_{\mu=1}^k \frac{1}{L_\mu} \geq 2k\, v^{1-\frac{1}{k}}\,\Sigma
\end{equation}
for the case of coordinate-aligned rectangular subsystems.

Hence, the lower bounds \eqref{sec:nonbinary-obc-a1} and \eqref{sec:nonbinary-obc-b1} for arbitrary-shape subsystems and the bounds \eqref{sec:nonbinary-obc-a2} and \eqref{sec:nonbinary-obc-b2} for coordinate-aligned rectangular subsystems from the OBC system are simply multiplied by a factor $2$ for PBC.

All resulting lower bounds for the TTNS exceed the MPS cost exponent $3\Sigma$ from Eq.~\eqref{eq:hypercube-MPS}, and even the cost exponent $6\Sigma$ for simulations with finite OBC MPS. For $z\geq 4$, both rectangular-subsystem TTNS bounds also exceed the exponent $8\Sigma$ for balanced binary TTNS in Eq.~\eqref{eq:hypercube-TTNS}. Note that $2(2+\sqrt{6})\approx 8.90$.

\section{Systems with log-area-law entanglement}\label{sec:logAreaLaw}
The analysis for states that obey a strict area law~\eqref{eq:areaLaw} in Secs.~\ref{sec:M-A}-\ref{sec:nonbinary} extends to states with a logarithmically enhanced area-law entanglement such as low-energy states of fermionic systems with a regular Fermi surface \cite{Gioev2005,Barthel2006-74,Li2006,Leschke2014-112}.

For a subsystem $\A_i$ with a characteristic linear size
$\ell_i$, the log-area law scaling means
\begin{equation}\label{eq:logAreaLaw}
	S_{\alpha,i} = b_\alpha|\partial\A_i|\ln\ell_i +o(|\partial\A_i|\ln\ell_i).
\end{equation}
As for the strict area law, a common coefficient $b_\alpha$ is assumed for the relevant interfaces. For free fermions, $b_\alpha$ is determined
by an integral over the subsystem boundary and the Fermi surface \cite{Gioev2005,Leschke2014-112}. Cubic symmetry makes its density equal for coordinate-aligned planar faces as employed in the MPS and TTNS subsystem splittings discussed in Secs.~\ref{sec:2D}-\ref{sec:hypercube}.

Let us consider scaling limits with fixed aspect ratios such that the macroscopic subsystems that determine the leading costs have $\ell_i=\kappa_i L$ with
constants $\kappa_i>0$ and $|\partial\A_i|\propto L^{D-1}$. Then, $\ln\ell_i=\ln L+\O(1)$, and Eq.~\eqref{eq:logAreaLaw} becomes
\begin{equation}\label{eq:logAreaLaw-L}
	S_{\alpha,i} = b_\alpha|\partial\A_i|\ln L + o\left(L^{D-1}\ln L\right).
\end{equation}
We assume that $b_\alpha$ is continuous at $\alpha=1$, with $b_1>0$. As an example, for free fermions, one has $b_\alpha=\frac{1+\alpha}{2\alpha}\,b_1$ \cite{Leschke2014-112}.

At fixed approximation accuracy $\delta$ and fixed R\'enyi parameters $\tilde\alpha>1>\alpha>0$, the additional terms in the logarithms of the bond-dimension bounds \eqref{eq:Mbound} are $\O(1)$ and $\O(\ln N)$, respectively. If $N$ grows at most polynomially with $L$, both are $o(L^{D-1}\ln L)$. The argument of Sec.~\ref{sec:c_alpha} that led to Eq.~\eqref{eq:M-A-precise}, taking the large-system limit before sending the R\'enyi parameters to $1$, therefore gives
\begin{equation}\label{eq:M-logAreaLaw}
	M_i=e^{b_1|\partial\A_i|\ln L+o(L^{D-1}\ln L)}.
\end{equation}

For the geometries that satisfy these assumptions, the cost comparison hence proceeds with the same geometric coefficients as for strict area laws, replacing $c_1L^{D-1}$ by $b_1L^{D-1}\ln L$. With the conventional dense tensor-operation costs considered in this work \cite{Note2}, we obtain
\begin{equation}
	\C_\MPS \sim e^{3 b_1L^{D-1}\ln L}\ \ \text{and} \ \
	\C_\TTNS\sim e^{m b_1L^{D-1}\ln L},
\end{equation}
where $m>3$ is the corresponding TTNS geometry factor. With $\ln(\C_\TTNS/\C_\MPS)=(m-3)b_1L^{D-1}\ln L+o(L^{D-1}\ln L)$, the asymptotic advantage of MPS therefore persists with a logarithmically enhanced exponent. Equivalently, the cost formulas from Secs.~\ref{sec:2D}-\ref{sec:nonbinary} can be read with the size-dependent parameter
\begin{equation}\label{eq:q-logAreaLaw}
	q(L)=e^{b_1\ln L}=L^{b_1}
\end{equation}
in place of the constant $q=e^{c_1}$ from Eq.~\eqref{eq:M-A}.

\newpage
\section{Conclusion}\label{sec:conclusion}
\begin{table}[b]
	\centering\renewcommand{\arraystretch}{1.5}
	\begin{tabular}{l @{\hskip 2ex} l @{\hskip 1.7ex} l  @{\hskip 1.4ex} l}
	  \toprule
	  Geometry & PBC & MPS cost & TTNS cost\\
	  \midrule
	  2D $L\times L$ square					& --	& $q^{3L}$	& $q^{4L\dots 4.25L}$ \\
	  2D $L\times L$ cylinder				& $y$	& $q^{3L}$	& $q^{5.5L}$ \\
	  2D $L_x\times L$ cylinder, $L_x\gg L$	& $y$	& $q^{3L}$	& $q^{8L}$ \\
	  2D $L\times L$ torus					& $y,x$	& $q^{3L*}$	& $q^{8L}$ \\
	  \midrule
	  3D $L\times L\times L$ cube			& --		& $q^{3L^2}$	& $q^{4L^2}$ \\
	  3D $L\times L\times L$ cube			& $z,y$		& $q^{3L^2}$	& $q^{5.5L^2}$ \\
	  3D $L_x\times L\times L$, $L_x\gg L$	& $z,y$		& $q^{3L^2}$	& $q^{8L^2}$ \\
	  3D $L\times L\times L$ cube			& $z,y,x$	& $q^{3L^2*}$	& $q^{8L^2}$ \\
	  \midrule
	  $D$-dimensional $L^D$ hypercube		& all		& $q^{3L^{D-1}*}$	& $q^{8L^{D-1}}$ \\
	  \bottomrule
	\end{tabular}
	\caption{\label{tab:summary}Summary of computation-cost scaling for simulations of different 2D, 3D, and higher-dimensional systems with MPS and balanced binary TTNS with rectangular subsystems under the assumption of entanglement \mbox{(log-)}area laws with parameter $q$ as defined in Eqs.~\eqref{eq:M-A} and \eqref{eq:q-logAreaLaw}, respectively. Asterisks indicate that, in MPS simulations of systems with PBC in the $x$ direction, the stated cost can be achieved as described in Ref.~\cite{Pippan2010-81} or by sending $L_x\to \infty$ and using infinite MPS, which is preferable in many applications. An alternative MPS algorithm for PBC with cost $\sim q^{5L^{D-1}}$ is described in Ref.~\cite{Verstraete2004-4}.}
\end{table}
Based on bounds for MPS and TTNS bond dimensions in terms of R\'{e}nyi entanglement entropies \cite{Barthel2026-9}, the determined scaling of computational costs as summarized in Table~\ref{tab:summary} and TTNS cost bounds in Sec.~\ref{sec:nonbinary} suggest that MPS are more efficient than TTNS for the simulation of large systems in $D\geq 2$ spatial dimensions with \mbox{(log-)}area-law entanglement. In terms of the system size, there is an exponential separation in the cost scaling, which is smallest for OBC and largest when applying PBC in all spatial directions.
Some remarks are in order:

While MPS are generally also considered favorable for 1D systems, TTNS have been useful in the study of disordered spin chains \cite{Goldsborough2014-89,Hikihara2023-5}, and the exponential separation derived for $D\geq2$ does not extend directly to \Emph{1D area-law systems}. In one dimension, interfaces are points such that, for fixed \emph{global} approximation accuracy $\delta$, the system-size-dependent factor in the upper bond-dimension bound \eqref{eq:Mbound} and various polynomial cost factors matter (see below).

In our discussion, internal-node TTNS tensors carried no physical indices for the single-site Hilbert spaces -- the physical indices were all assigned to leaf-node tensors. Depending on the specific scenario it can be beneficial to also \Emph{assign physical indices to internal nodes} \cite{Murg2010-82,Nakatani2013-138,Gunst2018-14}. Consider moving some of the physical indices from leaves to ancestor vertices with a bounded number of physical indices at each vertex. For any branch subsystem in a fixed number of top layers, only indices moved to ancestor vertices outside that branch are removed. Thus, each such subsystem loses at most $\O(1)$ sites, changing its R\'{e}nyi entanglement entropies by at most $\O(1)$. The additional physical indices also contribute only constant factors to the tensor-operation costs. Consequently, such reassignments leave the top-layer cost bounds unchanged and preserve the asymptotic separation between TTNS and MPS.

In the analysis, we have determined the leading exponential scaling of computational costs and suppressed \Emph{subexponential cost factors} according to Eq.~\eqref{eq:def-sim}. These include subleading terms in the entanglement \mbox{(log-)}area law as well as polynomial factors associated with (a) sums over Hamiltonian terms, (b) the total number of tensor operations per sweep, and (c) the explicit $N$-dependent factor in the upper bond-dimension bound \eqref{eq:Mbound} at fixed global approximation accuracy $\delta$ and R\'enyi parameter $\alpha$. Such subleading factors may give TTNS an advantage over MPS for small system sizes $L$ but, according to the asymptotic scaling, there should then exist a crossover point $L_\MPS$ beyond which MPS are more efficient.

The comparison in this work is based on tensor-operation costs per optimization or time-evolution step. In principle, the \Emph{required number of iterations} for Krylov subspace methods for the local energy optimization in DMRG \cite{White1992-11,White1993-10,McCulloch2008_04,Schollwoeck2011-326,Murg2010-82,Nakatani2013-138} and for local time-evolution problems in the time-dependent variational principle \cite{Haegeman2011-107,Vanderstraeten2019-7,Bauernfeind2020-8} as well as the total number of sweeps in groundstate optimization may need to be increased with increasing system size to reach a certain accuracy. However, these numbers are usually system-size independent or chosen to be system-size independent in practical applications. The analysis in this work would also apply if these numbers scaled polynomially.

In the cost scaling analysis, we have assumed that the target state obeys a \mbox{(log-)}area law. In principle, \Emph{intermediate states} might have a different entanglement scaling. However, we can generally assume that the simulation spends most time in a regime where the entanglement scaling of the target state applies.

Throughout the discussion, we have assumed conventional dense linear algebra operations, implying cubic costs for matrix multiplication, SVD, etc. \Emph{Fast linear-algebra methods} like Strassen's algorithm permit subcubic matrix multiplication and decompositions as well as corresponding speedups for tensor contractions \cite{Strassen1969-13,Demmel2007-108,Huang2016-690,Huang2018-40}. With such fast linear algebra, which is currently not widely used, the numerical cost exponents would change, but a simple lower bound from the TTNS tensor sizes establishes that the asymptotic advantage of MPS persists.

As mentioned in the introduction, TTNS and MPS with fixed bond dimensions do \emph{not} match the area-law entanglement structure of typical systems in $D\geq 2$ dimensions. \Emph{PEPS and MERA} \emph{do} reproduce area laws \cite{Verstraete2004-7,Verstraete2006-96,Barthel2010-105,Evenbly2014-89a}, but have much higher tensor-operation costs. Hence, there should be crossover points $L_\text{PEPS}$ and $L_\text{MERA}$, where PEPS and MERA become more efficient than MPS (and TTNS). A cost comparison between these methods is more challenging and model-specific, as they require different optimization techniques.

\Emph{Beyond applications in quantum physics}, MPS and TTNS are also employed for machine learning \cite{Cohen2016-29,Stoudenmire2016-29,Stoudenmire2018-3,Liu2019-21,Convy2022-3a,Chen2024-46}, the compression of deep neural networks \cite{Novikov2015-28,Yang2017-70,Yin2020_05}, the approximation of high-dimensional probability distributions \cite{Han2018-8,Barthel2018-97,Barthel2020-1,Crotti2023-120,Crotti2025-19}, and tensor completion in big-data analysis \cite{Bengua2017-26,Rauhut2015-419,DaSilva2015-481}.
In these contexts, MPS and TTNS are frequently referred to as tensor trains and the hierarchical Tucker format, respectively \cite{Oseledets2011-33,Hackbusch2009-15,Grasedyck2010-31}. Entanglement entropies are not immediately relevant in these cases and a cost comparison should be based on the scaling of measures like the mutual information, parameter redundancy, global self-similarity etc.

Finally, the presented analysis applies to the case of unconstrained tensors and would generally need adaptation for \Emph{tensor networks with tensor constraints} such as TTNS with CP-rank constraints \cite{Chen2022_05,Chen2024-46}.

\begin{acknowledgments}
I thank Elizabeth Bennewitz, Zohreh Davoudi, Alexey Gorshkov, Caroline Jang, Hersh Kumar, Alessio Lerose, Federica Surace, and Nikita Zemlevskiy for discussions that motivated this work as well as Simone Montangero, Luka Pave\v{s}i\'{c}, and Darvin Wanisch for helpful feedback.
\end{acknowledgments}

\appendix

\section{Derivation of TTNS tensor-operation costs}\label{appx:TTNScontractions}
Here, we derive the bond-dimension dependence of the TTNS costs in Eq.~\eqref{eq:cost-TTNS} for SVDs, effective-Hamiltonian applications (energy gradients), and the propagation of Hamiltonian environment tensors, considering conventional dense linear algebra \cite{Note2} and unconstrained TTNS tensors. As in Sec.~\ref{sec:contract}, we will often suppress factors and labels due to the single-site Hilbert-space dimensions $d_x$, sums over Hamiltonian terms, and Hamiltonian tensor-network bond dimensions $k_i$. Such factors are assumed to grow at most polynomially with system size, are subleading in light of the exponential scaling \eqref{eq:M-A} of the TTNS bond dimensions, and are covered by the subexponential terms included in the ``$\sim$'' notation of Eq.~\eqref{eq:def-sim}.

Let us first address binary TTNS (vertex degree $z=3$) in Secs.~\ref{appx:TTNScontractions-SVD}-\ref{appx:TTNScontractions-L4} and comment on TTNS with $z>3$ in Sec.~\ref{appx:TTNScontractions-z}.
Specifically, consider a $z=3$ TTNS tensor $C_x$ with bond dimensions $M_a,M_b,M_c$ and site dimension $d_x$ as shown in Fig.~\ref{fig:TTNS-operations}. For tensors without physical indices, $d_x=1$. At this vertex, we choose the labels such that
\begin{equation}\label{eq:appx-bond-order}
	M_a\geq M_b\geq M_c.
\end{equation}
This ordering is purely a local labeling convention. The computation cost \eqref{eq:cost-TTNS} can equivalently be written in the permutation-invariant form
\begin{equation}\label{eq:appx-cost-symmetric}
    \Theta\left(M_aM_bM_c\max\{M_a,M_b,M_c\}\right).
\end{equation}

The Schmidt rank across bond $i\in\{a,b,c\}$ cannot exceed the product of the other two bonds' Schmidt ranks and the site dimension $d_x$ at the vertex, such that we can assume
\begin{equation}\label{eq:appx-M-bound}
	M_i\leq d_x\prod_{j\neq i}M_j
\end{equation}
without loss of generality.

\subsection{Singular value decompositions}\label{appx:TTNScontractions-SVD}
Given the bound \eqref{eq:appx-M-bound} and the fact that a conventional SVD of a dense $M\times N$ matrix costs $\Theta(MN\min\{M,N\})$, the maximum SVD cost is incurred when moving the orthogonality center across bond $a$ as illustrated in Fig.~\ref{fig:TTNS-operations}b. Matricizing $C_x$ with dimensions $M_a\times(d_xM_bM_c)$, the SVD cost is
\begin{equation}\label{eq:appx-svdCost}
    \Theta(d_x M_a^2 M_b M_c),
\end{equation}
agreeing with Eq.~\eqref{eq:cost-TTNS}. An alternative QR decomposition has the same cost scaling.
A complete bidirectional sweep that moves the orthogonality center across every edge includes this direction. The other directions can be less expensive,
but their inclusion changes the total cost at the vertex by at most a constant factor.

\subsection{Energy gradients}\label{appx:TTNScontractions-G}
Consider the gradient
\begin{equation}\label{eq:appx-grad}
	\partial\bra\hat{H}\ket/\partial C_x^\dag
\end{equation}
of the energy expectation value with respect to the TTNS tensor $C_x$, where the Hamiltonian is encoded as a TTN operator with bond dimensions $k_i$ assumed to scale polynomially in the system size. A diagrammatic representation of the gradient is shown in Fig.~\ref{fig:TTNS-operations}d. To evaluate it, one employs the Hamiltonian environment tensors $\E_a$, $\E_b$, and $\E_c$ that represent the effective Hamiltonian on the associated branches of the tree obtained by cutting edges $a$, $b$, and $c$, respectively, and employing the corresponding reduced branch Hilbert spaces of dimension $M_i$. Thus, $\E_a$ is a $k_i\times M_i\times M_i$ tensor.

Suppressing Hamiltonian TTN operator tensors and bond indices, we can write the energy gradient~\eqref{eq:appx-grad} as the tensor
\begin{equation*}
	G_{\mu_a\mu_b\mu_c}
	=\sum_{\mu_a'\mu_b'\mu_c'}
	 (\E_a)_{\mu_a\mu_a'}(\E_b)_{\mu_b\mu_b'}(\E_c)_{\mu_c\mu_c'}(C_x)_{\mu_a'\mu_b'\mu_c'},
\end{equation*}
where $\mu_i,\mu_i'=1,\dotsc,M_i$ are the bond indices for edge $i$. According to Lemma~2 in Ref.~\cite{Barthel2025-111}, every (unconstrained) tensor network can be contracted optimally through pairwise contractions. Evaluating $G$ by contracting $C$ with $\E_a$, then contracting $\E_b$, and finally contracting $\E_c$ incurs the costs $\Theta(M_a^2M_bM_c)$, $\Theta(M_aM_b^2M_c)$, and $\Theta(M_aM_bM_c^2)$, respectively. Their sum
\begin{equation}\label{eq:appx-Heff-cost}
	\Theta\big(M_aM_bM_c(M_a+M_b+M_c)\big) = \Theta(M_a^2M_bM_c)
\end{equation}
is independent of the order in which the environment tensors are contracted with $C$, and it agrees with Eq.~\eqref{eq:cost-TTNS}.

\subsection{Propagation of Hamiltonian environments}\label{appx:TTNScontractions-E}
To propagate Hamiltonian environment tensors $\E_b$ and $\E_c$ through vertex $x$ toward
bond $a$, the relevant contraction, with Hamiltonian tensors and bond indices suppressed, is
\begin{equation*}
	(\tilde{\E}_a)_{\mu_a\mu_a'}
	=\!\!\!\sum_{\mu_b\mu_b'\mu_c\mu_c'}\!\!\!
	 (A_x)^*_{\mu_a\mu_b\mu_c}(\E_b)_{\mu_b\mu_b'}(\E_c)_{\mu_c\mu_c'}(A_x)_{\mu_a'\mu_b'\mu_c'}
\end{equation*}
as illustrated in Fig.~\ref{fig:TTNS-operations}c. Here, we have again suppressed Hamiltonian tensors, and the tensor $A_x$ has the same format as $C_x$ in the previous discussion. It arises from $C_x$ through the SVD that shifts the TTNS orthogonality center from vertex $x$ across bond $a$; see Fig.~\ref{fig:TTNS-operations}b.

In accordance with Eq.~\eqref{eq:cost-TTNS}, the contractions for $\tilde{\E}_a$ again incur costs \eqref{eq:appx-Heff-cost}: For example, contract $A_x$ with $\E_b$ at cost $\Theta(M_aM_b^2M_c)$, then contract $\E_c$ at cost $\Theta(M_aM_bM_c^2)$, and finally contract $A_x^\dag$ at cost $\Theta(M_a^2M_bM_c)$.
The same reasoning applies to propagation toward either of the other bonds, building $\tilde{\E}_b$ and $\tilde{\E}_c$ also at cost~\eqref{eq:cost-TTNS}.

\subsection{Relation to Lemma~4 in Ref.~\texorpdfstring{\cite{Barthel2025-111}}{Barthel 2025}}\label{appx:TTNScontractions-L4}
The relevant closed tensor network is the energy expectation value. Suppressing the Hamiltonian tensors and bond indices,
\begin{multline}
	\bra\hat{H}\ket=\sum_{\mu_a\mu_b\mu_c\mu_a'\mu_b'\mu_c'}
	(C_x)^*_{\mu_a\mu_b\mu_c}(C_x)_{\mu_a'\mu_b'\mu_c'}\\
	\times(\E_a)_{\mu_a\mu_a'}(\E_b)_{\mu_b\mu_b'}(\E_c)_{\mu_c\mu_c'}.
\end{multline}
Removing $C_x^\dag$ leaves the energy gradient $G_{\mu_a\mu_b\mu_c}$. Removing $\E_a$ and replacing $C_x,C_x^\dag$ by $A_x,A_x^\dag$
gives the propagated environment $\tilde{\E}_a$ and similarly for $\E_b$ and $\E_c$.

These are all single-tensor environments of the same closed network $\bra\hat{H}\ket$. Lemma~4 in Ref.~\cite{Barthel2025-111} states that their optimal
pairwise contraction sequences have the same leading cost scaling, as explicitly observed in Secs.~\ref{appx:TTNScontractions-G} and \ref{appx:TTNScontractions-E} and pointed out in Sec.~\ref{sec:contract}.

\subsection{General vertex degrees \texorpdfstring{$z$}{z}}\label{appx:TTNScontractions-z}
It is straightforward to extend the preceding arguments to TTNS with arbitrary vertex degrees $z\geq 2$.

Specifically, consider a TTNS tensor $(C_x)^{\sigma_x}_{\mu_1\ldots\mu_z}$ with $z$ bond indices of dimensions $M_i$ and a physical index
$\sigma=1,\dotsc,d_x$. In the following, we again suppress site-basis and Hamiltonian-dependent factors.

To move the orthogonality center from vertex $x$ across a bond $i$, one matricizes $C_x$ by separating $\mu_i$ from the remaining indices. An SVD or QR decomposition of the resulting $M_i\times\prod_{j\neq i}M_j$ matrix incurs the cost
\begin{equation}\label{eq:appx-svdCost-z}
	\Theta\Big(M_i\prod_{j=1}^z M_j\Big).
\end{equation}

As in Secs.~\ref{appx:TTNScontractions-G} and \ref{appx:TTNScontractions-E}, the energy gradient $\partial\bra\hat{H}\ket/\partial C_x^\dag$ and environment tensors $\tilde{\E}_i$ are contracted iteratively, which, in generalization of Eq.~\eqref{eq:appx-Heff-cost}, leads to costs
\begin{equation}\label{eq:appx-Heff-cost-z}
	\Theta\Big(\prod_{j=1}^z M_j\sum_i M_i\Big).
\end{equation}

Both Eqs.~\eqref{eq:appx-Heff-cost} and \eqref{eq:appx-Heff-cost-z} are in accordance with Eq.~\eqref{eq:cost-TTNS-z}.

\end{document}